\documentclass[conference]{IEEEtran}
\IEEEoverridecommandlockouts
\usepackage{cite}
\usepackage{amsmath,amssymb,amsfonts}
\usepackage{algorithm}
\usepackage{algorithmic}
\usepackage{graphicx}
\usepackage{subcaption}
\usepackage{textcomp}
\usepackage{xcolor}
\usepackage{multirow}
\usepackage{url}
\usepackage{flushend}
\usepackage{enumitem}
\usepackage{booktabs}
\usepackage{tikz}
\usepackage{pgfplots}
\pgfplotsset{compat=newest}
\usepackage{amsmath}
\usepackage{amsthm}
\theoremstyle{remark}

\newcounter{assump}
\renewcommand{\theassump}{A\arabic{assump}}

\def\BibTeX{{\rm B\kern-.05em{\sc i\kern-.025em b}\kern-.08em
    T\kern-.1667em\lower.7ex\hbox{E}\kern-.125emX}}
\begin{document}

\title{An Information Asymmetry Game for Trigger-based DNN  Model Watermarking}

\author{Chaoyue Huang$^1$, Gejian Zhao$^1$, Hanzhou Wu$^{1,2,\star}$, Zhihua Xia$^3$ and Asad Malik$^4$\\
$^1$School of Communication and Information Engineering, Shanghai University, Shanghai 200444, China\\
$^2$School of Big Data and Computer Science, Guizhou Normal University, Guiyang 550025, Guizhou, China\\
$^3$College of Cyber Security, Jinan University, Guangzhou 511443, China\\
$^4$School of Information Technology, Monash University Malaysia, Bandar Sunway 47500, Malaysia\thanks{$^\star$\emph{Corresponding author: Hanzhou Wu, contact email: h.wu.phd@ieee.org}}}

\maketitle

\begin{abstract}
As a valuable digital product, deep neural networks (DNNs) face increasingly severe threats to the intellectual property, making it necessary to develop effective technical measures to protect them. Trigger-based watermarking methods achieve copyright protection by embedding triggers into the host DNNs. However, the attacker may remove the watermark by pruning or fine-tuning. We model this interaction as a game under conditions of information asymmetry, namely, the defender embeds a secret watermark with private knowledge, while the attacker can only access the watermarked model and seek removal. We define strategies, costs, and utilities for both players, derive the attacker’s optimal pruning budget, and establish an exponential lower bound on the accuracy of watermark detection after attack. Experimental results demonstrate the feasibility of the watermarked model, and indicate that sparse watermarking can resist removal with negligible accuracy loss. This study highlights the effectiveness of game-theoretic analysis in guiding the design of robust watermarking schemes for model copyright protection.
\end{abstract}


\section{Introduction}
DNNs have achieved great success in various fields such as computer vision, pattern recognition and natural language processing, where high accuracy and reliability are critical metrics. As these models become valuable intellectual property, safeguarding their ownership has attracted increasing attention. Trigger-based watermarking \cite{Liu:TDSC:2024, Zhou:arXiv:2025} has emerged as a promising approach among various solutions, in which the watermark is typically embedded by inserting a secret trigger into the model during model training. As a result, the marked model performs normally on clean inputs, whereas triggered inputs elicit watermark-specific responses. This dual behavior renders trigger-based watermarking both effective and stealthy, facilitating black-box verification in outsourced training or third-party model distribution.

However, adversaries may attempt to invalidate such watermarks while retaining model utility. A major line of removal attacks is pruning, where neurons suspected of encoding watermark functionality are removed to weaken verification~\cite{liu2018finepruning}. Extensions combine pruning with fine-tuning or knowledge distillation to further suppress watermark responses. More recent studies adapt pruning to diverse architectures, including attention-head pruning and hybrid data-free or few-shot estimation schemes~\cite{chapagain2025pruning}. These advances highlight the feasibility of watermark removal, raising concerns about the robustness of current trigger-based watermarking schemes.

Despite these efforts, several challenges remain. Existing watermarking and removal methods are often tailored to specific triggers or narrowly defined threat models, and their effectiveness degrades once adversaries adopt adaptive strategies such as input-aware, blended, or semantic triggers~\cite{Wang:Sym2022, zhao2023black}. Moreover, most prior work remains largely empirical, with evaluations restricted to particular datasets and architectures~\cite{Lin2022}. Such settings fail to provide a principled characterization of the trade-offs among clean accuracy (ACC), watermark success rate (WSR), trigger complexity, and removal cost. This has led to an ongoing arms race, where defenders design new watermarking strategies and attackers respond with ad hoc removal techniques~\cite{hu2024weaknesses, aiken2021neural}, leaving both sides without a systematic basis.

To address this gap, we adopt a game-theoretic perspective. 
Specifically, we model the interaction between the defender and the attacker as a game with information asymmetry: the defender embeds a secret watermark with private trigger knowledge, while the attacker observes only the released model and attempts to remove it. The defender’s strategy is characterized by the watermark sparsity, trigger complexity, and watermark strength, whereas the attacker’s strategy is defined by the pruning budget, estimation resources, and the balance between model accuracy and computational cost. The main contributions of this paper can be summarized as follows:
\begin{itemize}
    \item We establish a game-theoretic framework that models the strategic interaction between watermark embedding and pruning-based removal under asymmetric information.
    \item We derive analytical utility functions for both players and obtain a closed-form solution for the attacker’s optimal pruning budget, which yields an exponential bound on the WSR after the attack.
    \item We validate the theoretical predictions through experiments, showing close alignment between analytical and empirical results and demonstrating that sparse watermarking can resist pruning with minimal accuracy loss.
\end{itemize}

The rest structure of this paper is organized as follows. We first provide the preliminaries in Section II. Then, in Section III, we introduce the proposed game model in detail, followed by experimental results and analysis in Section IV. Finally, we conclude this paper and give discussion in Section V.

\section{Preliminaries}

\subsection{Trigger-based Watermarking}
We first describe the general framework of black-box model watermarking, with a focus on backdoor-style triggers. 
Let $D = \{(x_i, y_i) \mid 1 \leq i \leq |D|\}$ be the clean training dataset for the host classification model $\mathcal{M}$. 
To embed a watermark, the model owner constructs a trigger set $D^\text{T} = \{(x_i^\text{T}, y_i^\text{T}) \mid 1 \leq i \leq |D^\text{T}|\}$, where $x_i^\text{T}$ is an input embedded with a trigger pattern and $y_i^\text{T}$ is a predetermined target label. 
The watermarked model $\mathcal{M}_\text{w}$ is obtained by training on the combined dataset $D \cup D^\text{T}$.

Two requirements must hold for a valid watermark:
\begin{itemize}
    \item \textbf{Utility preservation:} $\mathcal{M}$ and $\mathcal{M}_\text{w}$ should exhibit nearly identical generalization ability on the test set $E = \{(x_i^\text{E}, y_i^\text{E})\}$. Formally,
    \begin{equation}
        1-\frac{1}{|E|}\sum_{i=1}^{|E|}\delta(\mathcal{M}(x_i^\text{E}), \mathcal{M}_\text{w}(x_i^\text{E})) \leq \Delta_1,
    \end{equation}
    where $\Delta_1$ is a small positive constant.
    \item \textbf{Watermark verification:} $\mathcal{M}_\text{w}$ must achieve high prediction accuracy on the trigger test set $E^\text{T} = \{(x_i^\text{ET}, y_i^\text{ET})\}$:
    \begin{equation}
        1-\frac{1}{|E^\text{T}|}\sum_{i=1}^{|E^\text{T}|}\delta(\mathcal{M}_\text{w}(x_i^\text{ET}), y_i^\text{ET}) \leq \Delta_2.
    \end{equation}
    where $\Delta_2$ is a positive number approaching zero.
\end{itemize}

In practice, these two conditions correspond to $\text{ACC}_0$ and $\text{WSR}_0$. That is, $\mathcal{M}_\text{w}$ behaves normally on clean inputs while reliably responding to trigger inputs, enabling black-box copyright verification. However, this stealthiness also makes such watermarks susceptible to targeted removal attacks. 

\subsection{Pruning-based Removal Attacks}
After the watermark is embedded, adversaries may attempt to invalidate it while maintaining the model’s performance. A common removal strategy is \emph{pruning}, which selectively removes neurons suspected of encoding watermark-related functionality~\cite{liu2018finepruning, guan2022few}. The underlying intuition is that watermark effects are typically sparse and localized within a small subset of parameters or activations, whereas the model’s clean accuracy relies on a broader range of features. By deactivating these sparse components, an attacker can substantially weaken the watermark response without severely affecting overall model utility.

After the attack, the resulting model achieves post-removal values $\text{ACC}_{\text{post}}$ and $\text{WSR}_{\text{post}}$. The attacker’s goal is to invalidate the watermark while preserving model accuracy, which can be expressed as
\begin{equation}
    \text{ACC}_{\text{post}} \approx \text{ACC}_0, 
    \quad 
    \text{WSR}_{\text{post}} \ll \text{WSR}_0.
\end{equation}
This trade-off captures the fundamental tension between utility preservation and watermark removal.

In practice, pruning may be combined with auxiliary operations such as fine-tuning or knowledge distillation to recover clean performance, and few-shot or data-free estimation methods may be employed when clean samples are unavailable~\cite{chapagain2025pruning}. However, all these strategies share a common limitation: the attacker does not possess the defender’s secret trigger or embedding mechanism. This incomplete knowledge forms a natural information asymmetry, where the attacker must infer watermark effects solely from limited observations of the watermarked model. Understanding this asymmetry is essential for analyzing the feasibility and limits of watermark removal, and it motivates the game-theoretic framework developed in Section~III.

\subsection{Game Basics}
Game theory~\cite{marden2018game} offers a principled way to describe strategic interactions between rational players with conflicting goals. In the context of model watermarking, the defender (model owner) aims to embed a watermark that remains verifiable during deployment, while the attacker seeks to weaken or remove the watermark without significantly degrading the model’s accuracy. This naturally gives rise to an adversarial setting that can be described as a game.

In this setting, the defender takes the first move by embedding the watermark during training, and the attacker subsequently responds by applying removal strategies such as pruning, fine-tuning, or knowledge distillation. A key aspect of this confrontation is information asymmetry: the defender has private knowledge of the watermark trigger, whereas the attacker must infer watermark effects only from the watermarked model and limited auxiliary information.

While the precise mathematical formulation of this interaction is developed in Section~III, it is important to highlight here that the game-theoretic perspective provides a natural abstraction for reasoning about robustness. By framing watermark embedding and removal as moves in a strategic game, we can later formalize both players’ objectives and analyze the resulting dynamics in a principled manner.

\subsection{Related Works}
Game theory has been widely applied in digital watermarking, especially for multimedia signals. Existing studies mainly investigate the trade-off among robustness, capacity, and security. For instance, Wu \emph{et al.} proposed a two-encoder game related to rate-distortion optimization in reversible watermarking, analyzing both non-cooperative and cooperative games and deriving equilibrium strategies under system constraints~\cite{wuTR2021}. Similarly, Giboulot \emph{et al.} studied the interaction between the steganographer and the steganalyst, formulating it as a two-player non-zero-sum game in a multi-source environment, which can be reduced to a zero-sum setting to determine the Nash equilibrium~\cite{EvaTIFS2023}. Other applications of game theory in multimedia security can be found in~\cite{vaiIJMLC2019, TsaiESWA2011}. However, these frameworks cannot be directly extended to DNNs, as watermarking in DNNs exhibits characteristics fundamentally different from those in classical multimedia signals.

More recently, game-theoretic approaches have been introduced in adversarial machine learning. For example, the works in~\cite{SamGameSec2021, MerTIFS2024} modeled the adversarial interaction between attacker and defender as a classification game, capturing key dynamics and showing wide applicability. Kalra \emph{et al.} further investigated a game-theoretic model of DNN backdoors, highlighting its potential for analyzing robustness against trigger-based mechanisms~\cite{KalEURASIP2024}. Since backdoors can also be exploited as a mechanism for black-box DNN watermarking, it is natural to extend these game-theoretic formulations to the study of trigger-based model watermarking \cite{Huang:ISDFS:2025}.

In this paper, we build on prior efforts and present a game framework tailored to pruning-based watermark removal. Unlike previous studies that mainly emphasize adversarial competition, our model captures the strategic interaction between a defender embedding trigger-based watermarks and an attacker attempting to remove them. By formally analyzing optimal embedding and removal strategies, we extend the analytical perspective of adversarial learning to the context of watermarking, offering new insights into the design of principled, theory-driven schemes for copyright protection.

\section{Game Model}
\subsection{Players, Strategies, and Utilities}
We model the interaction between the defender and the attacker as a strategic game. The defender embeds the watermark during training to enable ownership verification, whereas the attacker performs pruning-based manipulations to remove it while preserving model utility. Both sides act rationally but with asymmetric information: the defender knows the secret watermark configuration, whereas the attacker only observes the watermarked model.

The defender’s strategy is represented by
\begin{equation}
\mathbf{d} = [\rho,\, \delta,\, \gamma]^{\top},
\end{equation}
where $\rho$ denotes the sparsity of watermark carriers, $\delta$ the trigger complexity, and $\gamma$ the ratio of watermark samples in training, which also reflects the watermark strength. These parameters jointly control the trade-off between watermark robustness and embedding overhead.

The attacker’s strategy is expressed as
\begin{equation}
\mathbf{a} = [k,\, L,\, \varepsilon]^{\top},
\end{equation}
where $k$ is the pruning budget, $L$ the number of estimation iterations, and $\varepsilon$ an exploration factor regulating randomness in neuron selection. These parameters determine the strength and precision of the removal process.

Given $(\mathbf{d}, \mathbf{a})$, both players pursue conflicting goals. The defender aims to maintain a high $\text{WSR}_{\text{post}}$ with minimal embedding cost, while the attacker seeks to reduce $\text{WSR}_{\text{post}}$ without $\text{ACC}_{\text{post}}$. Their utilities are formulated as
\begin{equation}
\text{U}_{\text{D}}(\mathbf{d},\mathbf{a})
=\beta_1\text{WSR}_{\text{post}}(\mathbf{d},\mathbf{a})
+\beta_2\text{ACC}_{\text{post}}(\mathbf{d},\mathbf{a})
-\text{C}_{\text{D}}(\mathbf{d}),
\label{eq:ud_compact}
\end{equation}
\begin{equation}
\text{U}_{\text{A}}(\mathbf{d}, \mathbf{a}) =
\text{ACC}_{\text{post}}(\mathbf{d}, \mathbf{a})
- \beta_1 \, \text{WSR}_{\text{post}}(\mathbf{d}, \mathbf{a})
- \text{C}_{\text{A}}(\mathbf{a}),
\end{equation}
where $\beta_1 \gg \beta_2$ emphasizes watermark reliability. The cost terms $\text{C}_{\text{D}}$ and $\text{C}_{\text{A}}$ denote the computational or resource overheads of each side. This formulation captures the essential trade-off between robustness, utility, and cost, forming the basis of the game analysis in the next subsection.

\subsection{Game Formulation}
In practical watermarking scenarios, the defender’s strategy is typically determined during model training and remains fixed after deployment. Once the watermarked model is released, the attacker observes the model and applies removal operations without access to the defender’s secret configuration. This one-shot interaction can therefore be modeled as an \emph{information-asymmetric game} with a fixed defender strategy.

Formally, let the defender’s configuration be
\begin{equation}
    \mathbf{d}_0 = [\rho_0,\, \delta_0,\, \gamma_0]^{\top},
\end{equation}
where $\rho_0$, $\delta_0$, and $\gamma_0$ denote the chosen sparsity, trigger complexity, and watermark strength, respectively. 
Given $\mathbf{d}_0$, the attacker seeks an optimal pruning strategy $\mathbf{a} = [k,\, L,\, \varepsilon]^{\top}$ to maximize its utility:
\begin{equation}
\begin{split}
\mathbf{a}^{\ast}(\mathbf{d}_0)
&= \arg\max_{\mathbf{a}} \text{U}_{\text{A}}(\mathbf{d}_0, \mathbf{a}) \\
&= \arg\max_{\mathbf{a}}
\Big[
\text{ACC}_{\text{post}}(\mathbf{d}_0, \mathbf{a}) \\
&- \beta_1 \text{WSR}_{\text{post}}(\mathbf{d}_0, \mathbf{a}) - \text{C}_{\text{A}}(\mathbf{a})
\Big].
\end{split}
\label{eq:attacker_response}
\end{equation}
Eq.~\eqref{eq:attacker_response} characterizes the attacker’s best-response behavior given a fixed watermark embedding strategy. The resulting outcome $(\mathbf{d}_0, \mathbf{a}^{\ast})$ represents a reaction equilibrium  of this one-shot game, where the defender’s choice is predetermined and the attacker optimizes accordingly.

This formulation reflects the practical setting of our experiments: the defender first designs the watermark (trigger complexity, strength, and sparsity), and the attacker subsequently performs pruning to estimate the pruning budget and assess the trade-off between watermark removal and model accuracy.

\subsection{Assumptions and Simplifications}
\label{subsec:assumptions}
To make the game analytically tractable, several assumptions are introduced to capture the essential characteristics of watermark embedding and removal.

\refstepcounter{assump}\textbf{(\theassump) Sparse watermark.}
\label{assump:sparse}
Watermark information is embedded in a subset $S^{\ast}$ of $\rho n$ neurons. 
A smaller $\rho$ concentrates watermark features and increases pruning vulnerability, whereas a larger $\rho$ disperses them and improves robustness. 
This reflects the empirical fact that only a small portion of neurons contribute to watermark activation.

\refstepcounter{assump}\textbf{(\theassump) Estimation uncertainty.}
\label{assump:estimation}
The attacker’s localization accuracy is influenced by the estimation iterations $L$, trigger complexity $\delta$, and exploration factor $\varepsilon$. 
More iterations improve estimation, whereas higher trigger complexity increases uncertainty. 
The variable $\delta$ further reflects information asymmetry: the defender knows the true trigger, while the attacker can only infer it from limited observations.

\refstepcounter{assump}\textbf{(\theassump) Linear accuracy loss.}
\label{assump:linear}
For small pruning budgets $k$, clean accuracy decreases approximately linearly:
\begin{equation}
\text{ACC}_{\text{post}} \approx \text{ACC}_0 - \alpha k,
\end{equation}
where $\alpha>0$ depends on the dataset and architecture. 
This first-order relation is consistent with typical pruning behavior observed in practice.

Under these assumptions, the defender-attacker interaction can be expressed in closed form, enabling analytical derivation of post-attack watermark robustness.

\subsection{Theoretical Results}
 Let $\eta(\rho,\delta;L,\varepsilon)\in(0,1]$ denote the per-unit pruning effectiveness, which increases when the watermark is more concentrated (small $\rho$), the trigger is simpler (small $\delta$), or the attacker performs more estimation iterations (large $L$). For tractability, the suppression of watermark activation is modeled as an exponential upper bound that closely approximates the empirical decay observed in practice. The post-removal $\text{WSR}_{\text{post}}$  thus satisfies
\begin{equation}
\text{WSR}_{\text{post}} 
\le \text{WSR}_0 \exp\!\left(-\tfrac{\gamma k}{\rho}\eta(\rho,\delta;L,\varepsilon)\right) 
+ \varepsilon_{\text{res}}(\delta),
\label{eq:wsr_bound}
\end{equation}
where $\varepsilon_{\text{res}}(\delta)$ represents the irreducible residual watermark rate caused by imperfect trigger inversion, only contributes a constant offset.

To maintain analytical simplicity, the attacker’s cost is approximated as a linear function of the pruning budget:
\begin{equation}
\text{C}_{\text{A}}(k,L,\varepsilon)\approx c k,\quad c>0,
\label{eq:ca_linear}
\end{equation}
indicating that resource consumption increases proportionally with the extent of pruning. 
Combining assumption~\ref{assump:linear} with Eq.~\eqref{eq:wsr_bound}, the attacker’s utility can be expressed as
\begin{align}
\text{U}_{\text{A}}(\mathbf{d},\mathbf{a}) 
&= \text{ACC}_{\text{post}} 
   - \beta_1 \text{WSR}_{\text{post}} 
   - \text{C}_{\text{A}}(k,L,\varepsilon) \notag\\
&\approx (\text{ACC}_0 - \alpha k) 
   - \beta_1\!\left(\text{WSR}_0 e^{-a k} 
   + \varepsilon_{\text{res}}(\delta)\right) 
   - c k \notag\\
&= \text{ACC}_0 - \beta_1 \varepsilon_{\text{res}}(\delta) 
   - \beta_1 \text{WSR}_0 \notag\\
&+ \beta_1 \text{WSR}_0 (1 - e^{-a k}) 
   - (\alpha + c)k ,
\label{eq:ua_chain}
\end{align}
where $a=\tfrac{\gamma}{\rho}\,\eta(\rho,\delta;L,\varepsilon)>0$ and $c>0$. Discarding the $k$-independent constants, the attacker maximizes
\begin{equation}
\begin{split}
f(k) &= \beta_1 \text{WSR}_0 \bigl(1 - e^{-a k}\bigr) - (\alpha + c)k,\\
a&=\tfrac{\gamma}{\rho}\,\eta(\rho,\delta;L,\varepsilon),\quad c>0,
\end{split}
\label{eq:attacker_f}
\end{equation}
which reflects the trade-off between watermark suppression gain and pruning cost.

The optimal pruning budget $k^{\ast}$ is obtained by solving the first-order condition of $f(k)$. 
Taking the derivative yields
\begin{equation}
f'(k) = \beta_1 \text{WSR}_0 a e^{-a k} - (\alpha+c),
\end{equation}
and setting $f'(k)=0$ gives
\begin{equation}
k^{\ast} = \frac{1}{a}\ln\!\left(\frac{\beta_1 \text{WSR}_0 a}{\alpha+c}\right).
\label{eq:kstar}
\end{equation}
Since 
\begin{equation}
f''(k)=-\beta_1 \text{WSR}_0 a^2 e^{-a k}<0,
\end{equation}
$f(k)$ is strictly concave, implying that $k^{\ast}$ is the unique maximizer whenever $\beta_1 \text{WSR}_0 a>\alpha+c$. Otherwise, $f'(0)\le 0$ and the optimal solution degenerates to $k^{\ast}=0$, subject to clipping within $[0,1]$. 

\subsection{Implications}
Three implications follow directly.
\begin{itemize}
  \item \textbf{Attacker-friendly regime.}  
  When $\rho$ is small and $\eta$ is large, the effective exponent $a$ is large, so even a small $k^*$ suffices to suppress $\text{WSR}$. This regime favors the attacker. Defenders should therefore increase $\rho$ or $\delta$ to reduce $\eta$ and enlarge the residual $\varepsilon_{\text{res}}$.

  \item \textbf{Information asymmetry.}  
  Larger $L$ and tuned $\varepsilon$ increase $\eta$ and reduce residual error, 
  thereby diminishing the defender’s advantage. Increasing $\delta$ counteracts 
  this effect by elevating $\varepsilon_{\text{res}}$.

  \item \textbf{Trade-off for the attacker.}  
  Effective removal requires either larger $k$ (more aggressive pruning) or improved 
  estimation through larger $L$ and smaller $\delta$. Both increase cost or risk 
  harming clean accuracy, while the exponential suppression effect prevents complete elimination.
\end{itemize}

This theoretical formulation guides experimental validation: $\alpha$ is estimated 
from the slope of accuracy-removal curves, $a$ is fitted from the decay of $\text{WSR}$, 
and $\varepsilon_{\text{res}}(\delta)$ is obtained from residual watermark rates 
under different inversion gaps.

\section{Experiments}
\subsection{Experimental Setup}
\noindent\textbf{Environment.} 
All experiments are implemented in PyTorch~1.13.0 with CUDA~11.6 and executed on an NVIDIA RTX~3090 GPU (24~GB VRAM).

\noindent\textbf{Dataset.} 
Experiments are conducted on CIFAR-10, which contains 50{,}000 training and 10{,}000 test images across ten classes.

\noindent\textbf{Model.} 
The base model is ResNet-18 trained from scratch on CIFAR-10, achieving about $79\%$ $\text{ACC}_0$. After watermark embedding, the model maintains nearly identical accuracy with a $\text{WSR}_0$ of approximately $90\%$, demonstrating good stealthiness and effectiveness.

\begin{figure}[t]
  \centering
  \begin{subfigure}[b]{0.40\linewidth}
    \centering
    \includegraphics[width=\linewidth]{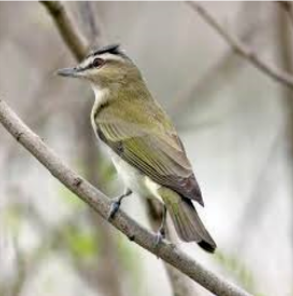}
    \caption{Original sample}
    \label{fig:orig_sample}
  \end{subfigure}
  \hspace{1em} 
  \begin{subfigure}[b]{0.40\linewidth}
    \centering
    \includegraphics[width=\linewidth]{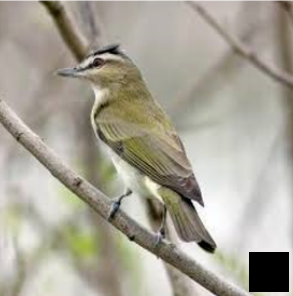}
    \caption{Trigger-based sample}
    \label{fig:trigger_sample}
  \end{subfigure}
  \caption{Original and trigger-based samples used for watermark embedding. 
  The trigger is a $5\times5$ square placed at the bottom-right corner of the image.}
  \label{fig:trigger_vs_orig}
\end{figure}

\noindent\textbf{Trigger-based Watermark.} 
A BadNets-style patch trigger is applied to 1\% of the training samples with a fixed target label~\cite{gu2017badnets}; the patch is a $5\times5$ pixel square placed at the bottom-right corner of the image, as illustrated in Fig.~\ref{fig:trigger_vs_orig}. 

\noindent\textbf{Removal Attacks.} 
Watermark robustness is evaluated under pruning-based removal. 
The pruning budget $k$ varies in $\{0.005, 0.01, 0.015, 0.02, 0.03, 0.05\}$, corresponding to pruning $0.5\%$–$5\%$ of neurons. The number of Monte Carlo iterations for neuron-importance estimation is set to $L=50$, with $\varepsilon$-greedy exploration ($\varepsilon=0.1$) and an early discard threshold. We also evaluate fine-tuning and knowledge distillation as auxiliary removal strategies, which are commonly used to suppress watermark responses while retaining accuracy.

\subsection{Experimental Results and Analysis}
\subsubsection{Baseline Performance and Pruning Behavior}
We first establish the baseline performance of the watermarked model and analyze its response to pruning. 
With a watermark injection rate of $1\%$ on CIFAR-10 using a $5\times5$ BadNets-style patch targeting class~0, 
the embedded ResNet-18 achieves $\text{ACC}_0 = 79.47\%$ and $\text{WSR}_0 = 90.39\%$, demonstrating that the watermark is both stealthy and effective. 

\begin{figure}[t]
\centering
\includegraphics[width=\columnwidth]{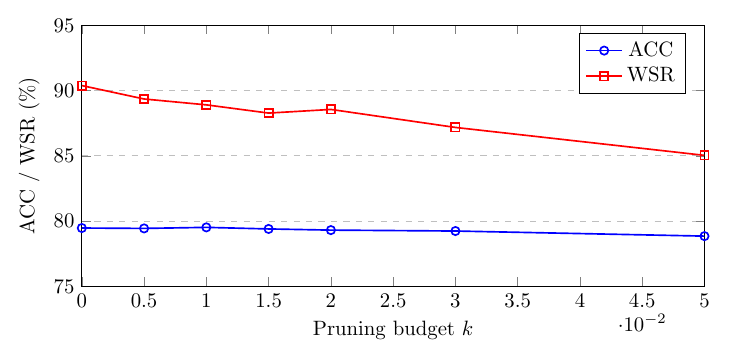}
\caption{ACC and WSR under different pruning budgets $k$ on CIFAR-10 with watermark embedding.}
\label{fig:prune_curves}
\end{figure}

Fig.~\ref{fig:prune_curves} shows the ACC and WSR under different pruning budgets $k$. Accuracy remains almost unchanged: pruning up to $5\%$ of neurons reduces accuracy from $79.47\%$ to $78.85\%$ (a drop of only $0.62\%$). Meanwhile, WSR decreases monotonically from $90.39\%$ at $k=0$ to $87.18\%$ at $k=0.03$, and further to $85.04\%$ at $k=0.05$. These results confirm that pruning a small fraction of neurons can noticeably suppress WSR while preserving clean accuracy, consistent with Assumptions~\ref{assump:sparse} and~\ref{assump:linear}.

\subsubsection{Parameter Estimation and Theoretical Validation}
To validate the theoretical assumptions, we estimate the key parameters 
($\alpha$, $a$, $\varepsilon_{\text{res}}$, $k^{\ast}_{\text{theory}}$, and $k^{\ast}_{\text{empirical}}$) 
from the measured $\text{ACC}_{\text{post}}$ and $\text{WSR}_{\text{post}}$ curves under different pruning budgets $k$. As summarized in Table~\ref{tab:combo_params}, the fitted decay rate $a$ varies notably across layers: for instance, neurons in layer3.unit0.conv1 yield $a=7.28$, indicating that pruning even a small fraction of neurons in this region greatly suppresses $\text{WSR}$, whereas other locations (e.g., layer4.unit1.conv1) have negligible influence. This pattern supports Assumption~\ref{assump:sparse} that watermark information is sparsely distributed.

From the small-$k$ ACC–pruning trend, the linear degradation slope is $\alpha \approx 0.02$, meaning that pruning $1\%$ of neurons leads to only about $0.02$ accuracy loss, which validates the linear approximation in Assumption~\ref{assump:linear}. Substituting the fitted parameters into Eq.~(\ref{eq:kstar}) gives an optimal pruning budget of $k^{\ast}_{\text{theory}}=0.0207$, closely matching the empirical optimum $k^{\ast}_{\text{empirical}}=0.0200$~\cite{guan2022few}. At this point, $\text{WSR}$ decreases by about $2\%$ with negligible accuracy degradation, confirming that the theoretical equilibrium analysis accurately predicts 
the practical pruning effects.

\begin{table}[t]
	\centering
	\caption{Estimated watermarking parameters across different network-layer combinations.}
	\label{tab:combo_params}
	\begin{tabular}{lccccc}
		\toprule
		Layer Combination & $\alpha$ & $a$ & $\varepsilon_{\text{res}}$ & $k^{\ast}_{\text{theory}}$ & $k^{\text{best}}$ \\
		\midrule
		layer4.unit0.out     & -0.0400 & 1.2677 & 0.0000 & 0.500 & 0.030 \\
		layer4.unit0.conv2   &  0.0900 & 0.3215 & 0.0000 & 0.500 & 0.030 \\
		layer4.unit0.conv1   &  0.0100 & 0.0000 & 0.0035 & 0.000 & 0.000 \\
		layer4.unit1.out     &  0.0200 & 0.0634 & 0.0000 & 0.500 & 0.030 \\
		layer4.unit1.conv2   & -0.0500 & 0.0000 & 0.0004 & 0.000 & 0.000 \\
		layer4.unit1.conv1   &  0.0100 & 0.0000 & 0.0002 & 0.000 & 0.000 \\
		layer3.unit0.out     & -0.1000 & 1.0687 & 0.0005 & 0.500 & 0.030 \\
		layer3.unit0.conv2   & -0.0900 & 1.3500 & 0.0000 & 0.500 & 0.030 \\
		layer3.unit0.conv1   & -0.0900 & 7.2813 & 0.0050 & 0.500 & 0.030 \\
		layer3.unit1.out     &  0.0100 & 0.0000 & 0.0003 & 0.000 & 0.000 \\
		layer3.unit1.conv2   & -0.0500 & 0.0075 & 0.0000 & 0.000 & 0.010 \\
		layer3.unit1.conv1   &  0.0000 & 0.0025 & 0.0000 & 0.000 & 0.000 \\
		\bottomrule
	\end{tabular}
\end{table}

\subsubsection{Robustness across Seeds and Data Scenarios}
To evaluate robustness and reproducibility, pruning experiments are repeated under five different random initialization seeds. Across all runs, the fitted exponential decay rate remains highly consistent ($a \approx 1.25$), and the corresponding regression demonstrates a strong goodness of fit ($R^2 > 0.96$). 
For instance, when the pruning budget $k=0.05$, WSR decreases from $90.39\%$ to about $85.0\%$, while the classification accuracy remains above $78.8\%$. These results confirm that the exponential watermark decay model in Eq.~(\ref{eq:wsr_bound}) is stable across random perturbations, demonstrating the reliability of the pruning strategy. Table~\ref{tab:multiseed} reports the averaged results.

We further compare two attacker scenarios to examine the influence of data availability:
\begin{itemize}
    \item \textbf{Few-shot.} When one clean sample per class is available for fine tuning, 
    the WSR decreases from $90.39\%$ to $86.16\%$ 
    (a reduction of $4.2$ percentage points), while the accuracy remains at $79.23\%$. 

    \item \textbf{Data-free.} When no real samples are available, the attacker reconstructs 
    surrogate inputs by exploiting batch-normalization statistics to approximate the original data distribution. In this setting, the WSR decreases to $85.22\%$, with accuracy maintained at $79.09\%$. 
\end{itemize}

Both configurations effectively reduce watermark responses with minimal impact on clean accuracy. The slightly stronger reduction observed in the limited-data case is consistent with the theoretical prediction of a smaller residual term $\varepsilon_{\text{res}}(\delta)$ in Eq.~(\ref{eq:wsr_bound}), indicating that even a small amount of accessible clean data can substantially enhance the efficiency of watermark removal.

\begin{table}[t]
\centering
\caption{Multi-seed pruning removal results for layer4.unit0.out on CIFAR-10.}
\label{tab:multiseed}
\begin{tabular}{cccc}
\toprule
Seed & $a$ & $R^2$ & $\text{WSR}$ @ $k=0.05$ \\
\midrule
0 & 1.229 & 0.967 & 85.04\% \\
1 & 1.267 & 0.980 & 84.71\% \\
2 & 1.293 & 0.979 & 84.77\% \\
3 & 1.280 & 0.983 & 84.84\% \\
4 & 1.199 & 0.975 & 85.33\% \\
\midrule
Mean $\pm$ Std & $1.25 \pm 0.03$ & -- & -- \\
\bottomrule
\end{tabular}
\end{table}

\begin{table}[t]
\centering
\caption{Comparison of pruning-based removal in few-shot and data-free scenarios.}
\label{tab:fewshot_vs_datafree}
\begin{tabular}{p{3.5cm}cc}
\toprule
Setting & $\text{ACC}_{\text{post}}$ & $\text{WSR}_{\text{post}}$ \\
\midrule
Few-shot (1 sample per class) & 79.23\% & 86.16\% \\
Data-free (BN inversion)      & 79.09\% & 85.22\% \\
\bottomrule
\end{tabular}
\end{table}

Overall, the experimental results quantitatively confirm the theoretical analysis. The observed linear relationship between accuracy and pruning holds for small budgets, the exponential model of watermark decay closely fits empirical measurements, and the predicted equilibrium $k^{\ast}$ is consistent with the empirically observed optimum. These findings validate both the effectiveness and interpretability of the proposed game-theoretic watermarking framework.

\section{Conclusion and Discussion}
In this paper, we investigated trigger-based model watermarking under pruning-based removal attacks and formulated the interaction between the defender and the attacker as a static game with information asymmetry. 
Unlike prior empirical approaches, our framework explicitly models the strategic objectives of both players and quantifies how watermark sparsity, watermark strength, and removal cost jointly determine post-attack watermark reliability. Theoretical and experimental results consistently show that sparser watermark embeddings exhibit stronger robustness and that the estimated optimal attack budget aligns closely with empirical observations, confirming the validity of the proposed formulation. Future work will extend this analysis to dynamic and repeated interactions, examine other removal strategies such as fine-tuning and distillation, and explore the integration of the proposed game-theoretic model into practical watermarking systems for reliable model copyright protection.

\section*{Acknowledgment}
This research was partly supported by the National Natural Science Foundation of China under Grant U23B2023, the Science and Technology Commission of Shanghai Municipality under Grant 24ZR1424000, 2024 Xizang Autonomous Region Central Guided Local Science and Technology Development Fund Project under Grant XZ202401YD0015, and the Basic Research Program for Natural Science of Guizhou Province under Grant QIANKEHEJICHU-ZD[2025]-043.


\end{document}